\documentclass[aps,amssymb,twocolumn]{revtex4}
\usepackage{graphicx}

\def\BSCCO{$\rm{Bi_2Sr_2CaCu_2O_{8+\delta}}$}
\def\LCO{$\rm{La_2CuO_4}$}
\def\LSCO{$\rm{La_{\it{x}}Sr_{1-\it{x}}CuO_4}$}
\def\LNSCO{$\rm{La_{1.6-\it{x}}Nd_{0.4}Sr_{\it{x}} CuO_4}$}
\def\YBCO{$\rm{YBa_2Cu_3O_{6.35}}$}
\def\rr{{\bf r}_1,{\bf r}_2}
\def\br#1{{\bf r}_{#1}}
\def\hl#1{\hat{Q}^{\alpha\beta}_{#1}}
\def\st#1{|#1\rangle}
\def\he#1{\hat{\bf e}_{#1}}
\def\Om{|\Omega\rangle}
\def\td{\hat{t}_{\alpha}^\dagger}
\def\tdi{\hat{t}_{\alpha,i}^\dagger}
\def\bd{\hat{b}^\dagger}
\def\nn{{\nonumber}}
\def\bA{{\bf A}}
\def \av#1{{\langle#1\rangle}}

\begin{document}

\title{Properties and Detection of Spin Nematic Order in Strongly Correlated Electron Systems}
\author{Daniel Podolsky}
\author{Eugene Demler}
\affiliation{Department of Physics, Harvard University, Cambridge MA 02138}

\begin{abstract}
A spin nematic is a state which breaks spin SU(2) symmetry while
preserving translational and time reversal symmetries.  Spin
nematic order can arise naturally from charge fluctuations of a
spin stripe state. Focusing on the possible existence of such a
state in strongly correlated electron systems, we build a nematic
wave function starting from a $t-J$ type model. The nematic is a
spin-two operator, and therefore does not couple directly to
neutrons. However, we show that neutron scattering and Knight
shift experiments can detect the spin anisotropy of electrons
moving in a nematic background.  We find the mean field phase
diagram for the nematic taking into account spin-orbit effects.
%We discuss the Goldstone modes corresponding to quantum spin
% nematic order.
\end{abstract}
\pacs{PACS numbers:}

\maketitle

\section{Introduction}
\label{sec:introduction}

When ordered, classical spin systems can arrange in a number of
patterns, including (anti)ferromagnetic, canted, and helical
structures. In addition to these, quantum mechanics allows the
formation of a wealth of magnetic phases for quantum spins not
available to their classical counterparts. Due to its quantum
numbers, detection of such order is often difficult:  For
instance, Nayak has considered a generalization of spin density
wave (SDW) order, in which spin triplet particle-hole pairs of
non-zero angular momentum condense with a modulated
density\cite{Nayak2000}.  These states are characterized by spin
currents rather than spin densities:  thus, they do not couple at
linear order to probes such as photons, neutrons, or nuclear
spins. Only at second order do these phases couple to conventional
probes, e.g. in two-magnon Raman scattering.  Despite the
challenges involved in their detection, subtle forms of magnetic
ordering such as these may be necessary to explain phenomena such
as the specific heat anomaly in the heavy-fermion compound
URu$_2$Si$_2$\cite{Chandra02}, and the pseudogap regime in the
cuprates\cite{chakravarty}.

% Detection of such order is made difficult
%by it's quantum numbers. For instance, the $d$ density wave
%phase...

%Many exotic magnetic phases arise naturally from fractionalization
%in strongly correlated electron systems. In one possible scheme of
%fractionalization, electrons with spin-1/2 and charge-1 split into
%fermionic spinless excitations with charge-1 (holons), and bosonic
%chargeless excitations of spin-1/2 (spinons). Then, condensation
%of the bosonic spinons leads to various possible magnetic phases,
%including some with topological order..

Promising materials for the observation of exotic magnetic phases
include systems with strong antiferromagnetic fluctuations such as
the heavy-fermion compounds, the organic superconductors, and the
cuprates. For instance, the cuprates in the absence of carrier
doping are antiferromagnetic Mott insulators at low temperatures.
As carriers are introduced through doping, the nature of the
magnetic order evolves until, for optimally doped and overdoped
samples, the system becomes a metallic paramagnet. In between
these two limits, the underdoped cuprates have been argued to have
spin glass\cite{keimer1991} and stripe phases\cite{tranquada99}.
The proximity between Mott insulator and superconducting phases in
the cuprates makes them ideal systems to study the hierarchy by
which the broken symmetry of Mott insulators is
restored~\cite{Sachdev2003,Sachdev2003b}.

In this paper we will explore the possibility of detection of spin
nematic order, a different quantum magnetic phase, in strongly
correlated electron systems\cite{andreevGrishchuk,GorkovSokol}.
Nematic order has been proposed as a state originating from charge
fluctuations of stripe order\cite{zaanenNussinov,nussinovZaanen}.
%Even in cases where there is no static
%spin stripe order due to fluctuations, there is evidence of
%pinning of fluctuating stripe order by impurities and vortices (REFS).
A spin stripe is a unidirectional collinear spin density wave
(SDW), and it consists of antiferromagnetically ordered domains
separated by anti-phase domain walls, across which the direction
of the staggered magnetization flips sign. The order parameter is
\begin{eqnarray*}
{\bf S}({\bf r})={\bf \Phi}e^{i{\bf K}_s\cdot{\bf r}}+{\bf
\Phi}^*e^{-i{\bf K}_s\cdot{\bf r}},
\end{eqnarray*}
with ${\bf K}_s=(\pi,\pi)+\vec{\delta}$ corresponding to
antiferromagnetically ordered stripes with period $2\pi/|\delta|$.
Here, the complex vector ${\bf \Phi}=e^{i \theta}{\bf n}$ takes
its value within the manifold of ground-states $S_1\times
S_2/Z_2$; the $Z_2$ quotient is necessary not to overcount
physical configurations, as the transformation $e^{i\theta} \to
-e^{i\theta}$, ${\bf n}\to -{\bf n}$ does not modify ${\bf \Phi}$.
The real vector ${\bf n}$ gives the direction of the staggered
magnetization in the middle of a domain, while the phase factor
$e^{i\theta}$ specifies the location of the domain walls.  A shift
in $\theta$ by $2\pi$ translates the system by the periodicity of
the SDW, and thus leaves the system invariant. Associated with
collinear SDW order is charge order due to the modulations in the
amplitude of the local spin magnetization\cite{EK}, which can be
described by a generalized CDW order parameter
\begin{eqnarray*}
\delta\rho({\bf r})=\varphi e^{2i{\bf K}_s\cdot{\bf r}}+\varphi^*e^{-2i{\bf K}_s\cdot{\bf r}},
\end{eqnarray*}
for some $SU(2)$ invariant observable $\rho$, not necessarily the
electron density.  In the stripe picture, this CDW usually arises
from the accumulation of holes at the domain walls.

Stripes were first observed in elastic neutron scattering
experiments on the spin-1 nickelate insulator
La$_{2-x}$Sr$_x$NiO$_4$, and coexistence of stripes with
superconductivity was first observed in underdoped
{\LNSCO}\cite{tranquada95}, where the unidirectional character was
demonstrated by transport\cite{uchida99} and
photoemission\cite{Zhou99} measurements. In {\YBCO} and {\LNSCO},
it is observed that SDW order is first destroyed by spin
fluctuations\cite{mook}. In this case, memory of the charge
modulation can remain, even after averaging over the spin
direction, resulting in a spin-invariant CDW phase, whose presence
may explain recent STM measurements in {\BSCCO}
\cite{davis1,kapitulnik1}.

\begin{figure}
\includegraphics[width=8cm]{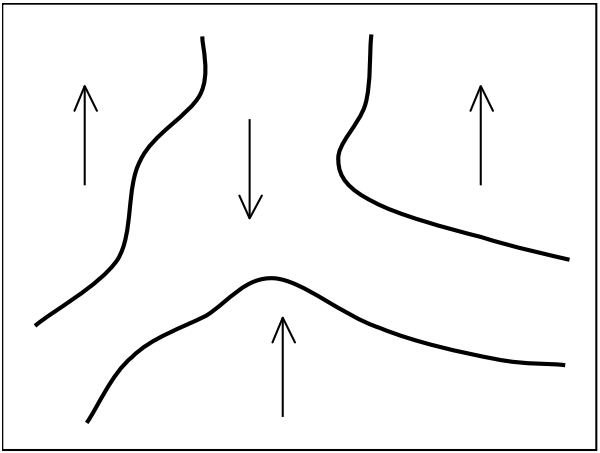}
\caption{Charge fluctuations of a spin stripe, provided that the
domain walls (curves) maintain their integrity. A spin nematic
state is a linear superposition of fluctuating domain
configurations such as these.  Within each domain, the staggered
magnetization (arrows) is well-defined, and it flips sign across
the anti-phase domain walls.  Due to fluctuations, translational
symmetry is restored to the system, and the magnetization has
expectation value zero at every point.  However, SU(2) symmetry is
not restored, as there is a preferred axis in which spins align,
modulo a sign. \cite{zaanenNussinov,nussinovZaanen}.
\label{fig:nemFluct}}
\end{figure}

For other materials, however, or in other regions of the phase
diagram, symmetry may be restored in a different order. In
particular, Zaanen and
Nussinov~\cite{zaanenNussinov,nussinovZaanen} proposed that many
experimental features of {\LSCO} can be explained by assuming that
the spin stripe order is destroyed by charge fluctuations.  In
this picture, dynamical oscillations in the anti-phase
domain-walls of a spin stripe lead to a restoration of
translational symmetry and a loss of N\'eel order.  However,
although both charge and spin order seem to be destroyed in this
process, $\delta\rho = 0$ and ${\bf S}=0$, full spin symmetry need
not be restored: So long as neither dislocations nor topological
excitations of the spin proliferate, the integrity of the domain
walls allows the staggered magnetization on each oscillating
domain to be well-defined. Thus, although the local magnetization
does not have an expectation value, the magnetization modulo an
overall sign does. This is nematic order, in which ${\bf S}$ and
$-{\bf S}$ are identified, see Fig.~\ref{fig:nemFluct}.  The order
parameter can be chosen to be $\langle S^{\alpha} S^{\beta}-S^2
\delta^{\alpha\beta}/3 \rangle\ \propto \langle n^{\alpha}
n^{\beta}-1/3 n^2 \delta^{\alpha\beta} \rangle\ne 0$, which has a
non-zero expectation value. Because translational invariance is
restored in this process, the nematic order parameter is spatially
uniform, instead of being modulated by some multiple of the SDW
wave vector. In addition, we expect the nematic to be uniaxial,
with a single preferred axis ${\bf n}$ (mod $Z_2$) inherited from
the nearby collinear SDW.

Direct observation of spin nematic order through conventional
probes is difficult.  For instance, neutrons do not couple to
nematic order, which is a spin-two operator.  Similarly, nematic
order is translationally invariant, and does not give Bragg peaks
in X-ray experiments. In principle, two-magnon Raman scattering
can probe nematic order, but in practice it is difficult to
separate the contribution due to the nematic from the creation of
two magnons\cite{zaanenNussinov}. Hence, although spin nematic
order can have important experimental consequences, e.g. for
antiferromagnetic correlations in magnetic field experiments on
superconducting samples \cite{zaanenNussinov}, its direct
detection remains a challenge.

The existence of other stripe liquid phases, different from the
spin nematic treated in this paper, as well as proposals for their
detection, are discussed in Ref.~\onlinecite{kivelson2003}. In
particular, the nematic state described there originates from
fluctuations of a unidirectional CDW that restore translational
invariance but, by maintaining a memory of the original
orientation of the CDW, break the rotational symmetry (point
group) of the lattice. Hence, unlike the spin nematic, this
``charge nematic" is SU(2) spin invariant, and only breaks a
discrete group.

\section{Spin nematic order parameter}
\label{sec:OP}

A spin nematic is a state that breaks spin SU(2) symmetry without
breaking time reversal invariance
\cite{andreevGrishchuk,gorkov1991}.  The presence of spin nematic
order can be observed in the equal time spin-spin correlator
\begin{eqnarray}
\langle \hat{S}^{\alpha}(\br{1})\hat{S}^{\beta}(\br{2})\rangle &=&
C(\rr)\delta^{\alpha\beta}+ \epsilon^{\alpha\beta\gamma}A^{\gamma}(\rr)\nn\\
&\,&+ Q^{\alpha\beta}(\rr). \label{spinDecomp}
\end{eqnarray}
This expression corresponds to the SU(2) decomposition
$(1)\otimes(1)\sim(0)_{\rm sym}\oplus(1)_{\rm asym}\oplus(2)_{\rm
sym}$. We consider the three terms appearing in eq.
(\ref{spinDecomp}) in turn. The scalar function $C$ is explicitly
spin invariant.  It contains important information regarding
charge order, but it does not help us in defining a nematic phase.
On the other hand, the pseudovector function ${\bf
A}=\langle\hat{\bf S}_1\times\hat{\bf S}_2\rangle$ gives a measure
of the non-collinearity of the spin vector field. However, in the
case at hand where the nematic state originates from charge
fluctuations of a collinear SDW phase, we expect $\bA$ to vanish.
This is supported by the fact that, from the point of view of the
Ginzburg-Landau free energy, the pseudovector ${\bf A}$ cannot
couple linearly to any function of the SDW order parameter ${\bf
\Phi}.$  As an aside, we note that for systems with spin exchange
anisotropy of the Dzyaloshinskii-Moriya (DM) form, the DM vector
will couple linearly to ${\bf A}$, as expected from the weak
non-collinearity (canting) in such systems. However, the
expectation value of $\bA$ in this case comes from explicit
breaking of the spin symmetry.

Thus, all information of interest to us is contained in the
symmetric spin-2 tensor $Q^{\alpha\beta}$.  We define a
symmetrized traceless spin correlator $\hat{Q}^{\alpha\beta}$,
\begin{eqnarray}
\hat{Q}^{\alpha\beta}(\rr)=\frac{1}{2}(\hat{S}^{\alpha}_1
\hat{S}^{\beta}_2+\hat{S}^{\beta}_1 \hat{S}^{\alpha}_2)-
\frac{\delta^{\alpha\beta}}{3}\hat{\bf S}_1\cdot\hat{\bf S}_2
\end{eqnarray}
whose expectation value yields $Q^{\alpha\beta}$ directly,
\begin{equation}
Q^{\alpha\beta}(\rr)=\langle \hat{Q}^{\alpha\beta}(\rr)\rangle.
\label{link}
\end{equation}
Starting from a SDW state, as domain wall fluctuations grow to
destroy charge order, translational invariance is restored to the
system, insuring that $Q^{\alpha\beta}(\rr)$ is independent of the
center-of-mass coordinate, {\it i.e.}
$Q^{\alpha\beta}(\rr)=Q^{\alpha\beta}({\bf r_1+R},{\bf r_2+R})$
for any displacement ${\bf R}$. This fact alone signals the
breaking of spin symmetry, since the choice of a non-trivial ({\it
i.e.} not proportional to $\delta^{ab}$) tensor $Q^{\alpha\beta}$
has been made across the system. Thus, while
$Q^{\alpha\beta}(\rr)$ decays exponentially with distance
$|\br{1}-\br{2}|$ due to the absence of long-range N\'eel order,
the onset of nematic order is reflected in the
translationally-invariant expectation value in the matrix-valued
function (\ref{link}).   Since the original SDW state has a single
preferred spin direction $\hat{\bf n}$, the ensuing nematic order
will be uniaxial,
\begin{eqnarray}
\begin{array}{c}
Q^{\alpha\beta}(\rr)= f(\br{1}-\br{2})Q_0^{\alpha\beta},\\ \\
Q_0^{\alpha\beta}=(n^{\alpha} n^{\beta}-\delta^{\alpha\beta}/3)S.
\end{array}
\label{nemfact}
\end{eqnarray}
Here, we have decomposed the order parameter into three component
objects: a function $f$ describing the internal structure of the
nematic; the unit vector director field ${\bf n}$; and the scalar
magnitude $S$.  As seen explicitly from (\ref{link}), the function
$f$ is parity-symmetric, $f({\bf r})=f(-{\bf r})$.  As will be
shown below, $f$ is dominated by the short range antiferromagnetic
correlations between spins, and therefore has a large contribution
at the wave vector $(\pi,\pi)$.  By definition, we choose this
contribution to be positive, $f(\pi,\pi)>0$.  With this
convention, $S>0$ corresponds to a N\'eel vector that is locally
aligned or anti-aligned with the director field ${\bf n}$
(sometimes referred to as the N$^+$ phase in the literature of
classical liquid crystals, see e.g.
Ref.~\onlinecite{deGennes1993}), while $S<0$ corresponds to a
N\'eel vector that is predominantly perpendicular to ${\bf n}$
(the N$^-$ phase).  We will show that, at low temperatures, $S>0$
due to the local antiferromagnetic correlations whereas, for
anisotropic systems at high temperatures, a phase with $S<0$ is
possible. Finally, we note that the unidirectional SDW state
(stripe phase) breaks the discrete rotational symmetry of a
tetragonal lattice.  This symmetry may be restored when charge
fluctuations destroy the SDW state to form the nematic.  In this
chase $f({\bf r})$ will be symmetric under rotations on the plane
by $\pi/2$, ${\bf r}\to R_{\pi/2}{\bf r}$. However, if a memory of
the orientation of the stripes survives the domain wall
fluctuations, $f$ will not have such symmetry, yielding a
``nematic spin-nematic'', {\it i.e.} a translationally invariant
system that is anisotropic in real space and in spin space.

Sections~\ref{sec:waveFn} and~\ref{sec:meanField} will be devoted
to understanding the behavior of the three component fields of the
nematic: $f$, ${\bf n}$, and $S$.  Then, in
Section~\ref{sec:detection} we will explore how this detailed
knowledge can be used in experimental searches for nematic order.

\section{Nematic wave function}
\label{sec:waveFn}

In order to explore the possible symmetry properties of the
function $f$, we study its short wave length structure by explicit
construction of a nematic operator on a small cluster.  As is well
known, it is impossible to describe nematic order in terms of a
single spin-$1/2$ particle, as the identification of ``up'' and
``down'' results in a trivial Hilbert space for the spin degree of
freedom. Another way to see this is that a spin-2 operator has a
vanishing expectation value with respect to any spin-$1/2$ state,
resulting in the identity $\hat{Q}^{\alpha\beta}_{ij}\equiv 0$
whenever $i=j.$

This limitation can be overcome by coarse-graining a group of
spins and constructing a nematic wave function out of them. In
order to preserve the underlying rotational symmetry of the
system, we carry this out on square $2\times2$ plaquettes of
spins, and  use energetic considerations to find the states most
likely to contribute to magnetic order. We take the $t$-$J$ model
as a starting point, analyzing the low energy Hilbert space in a
manner similar to the projected SO(5) approach of Zhang {\it et
al.} \cite{Zhang1999} or the CORE approach of Altman and Auerbach
\cite{Altman2002}. In these analyses, the lattice is first split
into plaquettes. On each plaquette, the Hamiltonian is
diagonalized, and the $m$ lowest energy states $\st{\psi_\nu}_i$
are kept, where $\nu\in\{1,\ldots, m\}$ and $i$ labels the
plaquette.  For a lattice composed of $N$ plaquettes, this allows
one to define a projected subspace ${\cal M}$ of the Hilbert space
spanned by $m^N$ factorizable states of the form
$\st{\psi_{\nu_1}}_1\otimes\st{\psi_{\nu_2}}_2\cdots\st{\psi_{\nu_N}}_N$.
%The CORE approach is completed by specifying a procedure to develop an effective
%low energy Hamiltonian acting on ${\cal M}$.
We assume %\cite{overlap}
that the ground state is well-contained in ${\cal M}$.

\begin{figure}
\includegraphics[width=8cm]{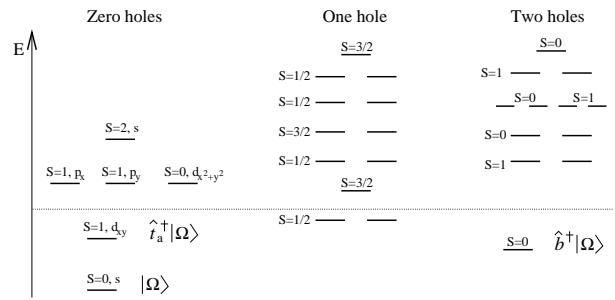}
\caption{ Spectrum of $t$-$J$ model on a $2\times 2$ plaquette.
Adapted from \cite{Altman2002}. \label{spect}}
\end{figure}

Figure \ref{spect} reproduces results in \cite{Altman2002} for a
$t$-$J$ model on a $2\times2$ plaquette.  The lowest energy
bosonic states are, at half filling, the $S=0$ ground state $\Om$
and the $S=1$ magnon triplet $\td\Om$, and for a plaquette with
two holes, an $S=0$ state $\bd\Om$.  In addition, there are two
low-lying $S=1/2$ fermion doublets with one hole; however, unbound
holes are dynamically suppressed, as supported by DMRG
calculations on larger lattices, and we shall exclude the one-hole
sector in what follows.  Then, to lowest order in the analysis, we
only keep the low-lying bosonic states, in terms of which a
general low-energy plaquette state can be written as
\begin{eqnarray}
|\psi\rangle_i=\left(s+m^{\alpha}\tdi+c\bd_i\right)\Om_i.
\label{lowplaq}
\end{eqnarray}

It is useful to get some intuition regarding the wave functions (\ref{lowplaq}).  Introducing
the total spin and staggered spin operators on a plaquette,
$\hat{\bf S}=\hat{\bf S}_1+\hat{\bf S}_2+\hat{\bf S}_3+\hat{\bf S}_4$
and $\hat{\bf N}=\hat{\bf S}_1-\hat{\bf S}_2+\hat{\bf S}_3-\hat{\bf S}_4$,
as well as the DM-type vector,
$\hat{\bf D}=\hat{\bf S}_1\times\hat{\bf S}_2-\hat{\bf S}_2\times\hat{\bf S}_3+
\hat{\bf S}_3\times\hat{\bf S}_4-\hat{\bf S}_4\times\hat{\bf S}_1$,
we find
\begin{eqnarray}
\hat{S}_{\alpha} \sim i \epsilon_{\alpha\beta\gamma}\hat{t}^\dagger_{\beta} \hat{t}_\gamma,\nn\\
\hat{N}_{\alpha} \sim \hat{t}_{\alpha}+\hat{t}^\dagger_{\alpha},\nn\\
\hat{D}_{\alpha} \sim
i(\hat{t}_{\alpha}-\hat{t}^\dagger_{\alpha}),\nn
\end{eqnarray}
%\begin{eqnarray}
%\hat{N}_{\alpha} |\psi\rangle_i \sim (s \tdi+m_{\alpha})\Om_i,\\
%\hat{S}_{\alpha} |\psi\rangle_i \sim i \epsilon_{abc}m_b \hat{t}_{c,i}^\dagger\Om_i,
%\end{eqnarray}
up to positive multiplicative factors and up to terms lying outside of the projected low-energy space.  In particular, we
find the expectation values on a single plaquette
\begin{eqnarray}
\langle\hat{\bf S}_i\rangle \sim {\bf m}^*\times{\bf m},\nn\\
\langle\hat{\bf N}_i\rangle \sim {\rm Re} (s^*{\bf m}),\nn\\
\langle\hat{\bf D}_i\rangle \sim {\rm Im} (s^*{\bf m}),\nn\\
\langle \hat{b}_i\rangle \sim s^* c,\nn\\
\langle \hat{\bf N}_i\hat{b}_i\rangle \sim {\bf m}^* c.
\label{expect}
\end{eqnarray}
The last two expressions describe local singlet and triplet superconductivity, respectively.

One can build a nematic on a cluster of plaquettes by choosing a
factorizable state, in which $s=0$ and ${\bf m}$ is a constant
real vector on every plaquette,
\begin{eqnarray}
|\psi_{\rm fact}\rangle=\prod_i(m^{\alpha}
\tdi+c\bd_i)\Om_i\label{factorizable}
\end{eqnarray}
The constraint that ${\bf m}$ is real insures that no long range
ferromagnetic or N\'eel orders develop, whereas nematic order does
due to the SU(2)-symmetry breaking choice of the vector ${\bf m}$.
Note from (\ref{expect}) that $|\psi_{\rm fact}\rangle$ is also a
triplet superconducting state (except at half-filling, where
$c=0$).  Order of this type is found, for instance, in the triplet
superconducting state of quasi-one dimensional Bechgaard salts,
where the triplet superconducting order parameter is constant
along the Fermi surface due to the splitting of the Fermi surface
into two disjoint Fermi sheets\cite{Podolsky2004}.

On the other hand, in applications to materials such as the high
$T_c$ cuprates, we would like to introduce a nematic state that is
a singlet superconductor, instead of triplet. For this, the use of
non-factorizable states is necessary. For instance, introducing a
local angle variable $\theta_i\in [0,2\pi)$ on each plaquette,
consider the state
\begin{eqnarray}
|\psi_1\rangle&=& \int{
(d\theta_1\,d\theta_2\ldots)}\label{nonfact}
\\&\,&\left[\prod_i(s+m^{\alpha}\cos({\bf
Q}\cdot r_i-\theta_i) \tdi+c\bd_i)\Om_i\right], \nn
\end{eqnarray}
where ${\bf m}$ is a constant real vector, and ${\bf Q}$ is the
wave vector of the underlying SDW.   If the angle $\theta_i$ were
held constant across the lattice, long range SDW order would
ensue.  In contrast, by integrating independently over the
$\theta_i$ at different sites, we introduce charge fluctuations
that average out the local magnetization to zero. Hence, the state
(\ref{nonfact}) has restored translational and time-reversal
symmetry, with only singlet superconductivity and nematic order
surviving.  We note that (\ref{nonfact}) ignores correlations
between spin degrees of freedom, controlled by
$\hat{t}_{\alpha}^\dagger$, and charge degrees of freedom,
controlled by $\hat{b}^\dagger$. More complex nematic wave
functions that take this effect into account are given in
Appendix~\ref{App:spinCharge}. On the other hand, in practical
calculations, one may consider a slightly simpler wave function
than (\ref{nonfact}) by replacing the $\theta_i$ by local Ising
variables $\sigma_i=\pm 1$ on each plaquette.  This leads to the
wave function
\begin{eqnarray}
|\psi_2\rangle=\sum_{\{ \sigma_1,\sigma_2\ldots
\}=\pm1}\left[\prod_i(s+\sigma_i
m^{\alpha}\tdi+c\bd_i)\Om_i\right]. \label{nonfact2}
\end{eqnarray}
As before, this state has time-reversal and translational
symmetries restored, with only spin nematic and superconducting
orders surviving.  Finally, note that in order to produce a
nematic state without any type of superconductivity (singlet or
triplet), it is necessary to introduce {\it another} fluctuating
Ising variable which multiplies the term $c\bd_i$ in
(\ref{nonfact2}), thus randomizing the relative phase between all
three components of the wave function.

\begin{figure}
\includegraphics[width=8cm]{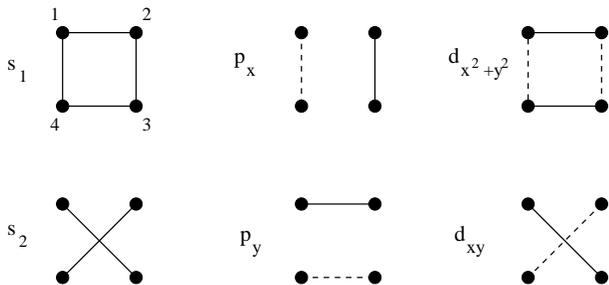}
\caption{ Basis of plaquette nematic operators with well-defined
symmetry properties. The solid lines denote links with weight
$+1$, dashed lines have weight $-1$. Thus, for instance,
$\hat{P}_{s_1}=(\hat{Q}_{12}+\hat{Q}_{23}+\hat{Q}_{34}+\hat{Q}_{41})/2$,
which in terms of the $\alpha_{ij}$ vectors reads ${\bf
\alpha}^{s_1}=(\he{12}+\he{23}+\he{34}+\he{41})/2$.  The other
basis vectors are ${\bf \alpha}^{s_2}=(\he{13}+\he{24})/\sqrt{2}$,
${\bf \alpha}^{p_x}=(\he{23}-\he{41})/\sqrt{2}$, ${\bf
\alpha}^{p_y}=(\he{12}-\he{34})/\sqrt{2}$, ${\bf
\alpha}^{d_{x^2+y^2}}=(\he{12}-\he{23}+\he{34}-\he{41})/2$, and
${\bf \alpha}^{d_{xy}}=(\he{13}-\he{24})/\sqrt{2}$.
\label{symplaq}}
\end{figure}

Despite the fact that we have considered many different wave
functions, depending on the possible coexistence of nematic order
with different types of superconductivity, the analysis above
allows us to make strong predictions on the dominant short wave
length dependence of $f$. This is because, of all the low energy
plaquette states kept in the projected Hilbert space, only the
triplet $\td\Om$ breaks SU(2) symmetry. Hence, only this state can
contribute directly to the nematic order parameter. There are six
different pairs $i\ne j$ of sites on a $2\times 2$ plaquette, and
the most general spin-2 operator $\hat{P}^{\alpha\beta}$ on a
plaquette can be written as a linear combination of the six
``link'' operators $\hat{Q}_{ij}$,
\begin{eqnarray}
\hat{P}^{\alpha\beta}=\sum_{\{ i\ne j \} \in
\Box}\alpha_{ij}\hat{Q}^{\alpha\beta}_{ij}. \label{plaqnemop}
\end{eqnarray}
It is useful to introduce an inner product for real functions
$\alpha_{ij}$ on the links of a plaquette, according to
$(\alpha,\beta)=\sum_{\{i\ne j\}\in \Box}\alpha_{ij}\beta_{ij}$.
With this, we can write a normalized basis of plaquette operators
$\hat{P}_\eta$, corresponding to 6 linearly independent functions
$\alpha^\eta_{ij}$ which satisfy
$(\alpha^\eta,\alpha^\nu)=\delta^{\eta\nu}$. We choose a basis of
eigenoperators of the symmetries of the plaquette, composed of
operators with $p_x$, $p_y$, $d_{x^2+y^2}$ and $d_{xy}$
symmetries, see Figure \ref{symplaq}, in addition to two operators
with $s$-wave symmetry,
\begin{eqnarray}
\begin{array}{c}
\hat{P}^{\alpha\beta}_{s_1}=\frac{1}{2}\left(\hl{12}+\hl{23}+\hl{34}+\hl{41}\right),
\\ \\
\hat{P}^{\alpha\beta}_{s_2}=\frac{1}{\sqrt{2}}\left(\hl{13}+\hl{24}\right).
\end{array}
\end{eqnarray}

It is easy to see that among the six operators (\ref{plaqnemop}),
only those with $s$-wave symmetry get a non-vanishing expectation
value with respect to the state (\ref{nonfact}).  What's more, the
linear combination
$\hat{P}_{s_F}=(\hat{P}_{s_1}+\sqrt{2}\hat{P}_{s_2})/\sqrt{3}$ has
a vanishing expectation value, so that it is possible to write an
orthonormal basis of operators in which only the basis operator,
\begin{eqnarray}
\hat{P}_{s_A}=\frac{1}{\sqrt{3}}(-\sqrt{2}\hat{P}_{s_1}+\hat{P}_{s_2}),
\end{eqnarray}
has a non-zero expectation value.  This can be understood as a
consequence of local antiferromagnetic correlations, since
$\hat{P}^{\alpha\beta}_{s_A}$ can be expressed in terms of the
plaquette staggered magnetization ${\bf N}$ as
\begin{eqnarray}
\hat{P}^{\alpha\beta}_{s_A}=\frac{1}{2\sqrt{6}}\left(N^{\alpha}N^{\beta}-{\bf
N}^2\frac{\delta^{\alpha\beta}}{3}\right).
\end{eqnarray}
Using the relation
\begin{eqnarray}
\hat{Q}_{ij}=\sum_\eta \alpha^{\eta}_{ij}\hat{P}_\eta
\end{eqnarray}
which holds for $i,j$ on the same plaquette, we see that
the dominant short-range contribution to the nematic order parameter is
\begin{eqnarray}
Q^{\alpha\beta}_{ij}=\alpha^{s_A}_{ij}\langle\hat{P}^{\alpha\beta}_{s_A}\rangle.
\end{eqnarray}
This is of the form (\ref{nemfact}) with
\begin{eqnarray}
 f({\bf
k})&=& \alpha^{s_A}({\bf
k})\nonumber\\
&=&\frac{4}{\sqrt{6}}(\cos k_x \cos k_y-\cos
k_x-\cos
k_y), \label{domform}\\
Q_0^{\alpha\beta}&=&
\langle\hat{P}^{\alpha\beta}_{s_A}\rangle.\nonumber
\end{eqnarray}
Note that we can relate the real vector ${\bf m}$ in Eqs.
(\ref{nonfact}) and (\ref{nonfact2}) to the director field ${\bf
n}$ of the nematic, appearing in (\ref{nemfact}), through
\begin{eqnarray}
\langle \hat{P}^{\alpha\beta}_{s_A}\rangle\propto m^{\alpha}
m^{\beta}-m^2\delta^{\alpha\beta}/3,
\end{eqnarray}
from which we conclude that ${\bf n}\propto{\bf m}$.  This can be
used to constrain the sign of $S$,
\begin{eqnarray}
S&=& \frac{3}{2}n^{\alpha}n^{\beta}\langle
\hat{P}^{\alpha\beta}_{s_A}\rangle\nonumber>0.
\end{eqnarray}
Therefore, at low temperatures, where the state of the system is
dominated by the low energy plaquette states, $S$ is positive.
%{\bf Another possible interpretation is in terms of collinear vs non-collinear order:
%a real $m$ is likely to be associated with collinear order.  We could then argue that
%interactions at slightly higher wave lengths favor collinearity (and simultaneously local
%antiferromagnetism?). This then supports that $\gamma>0$ in the next section,
%implying that $n$ lies along the easy plane, giving a very strong experimental signature}.

\section{Mean field analysis}
\label{sec:meanField}

\begin{figure}
\includegraphics[width=6cm]{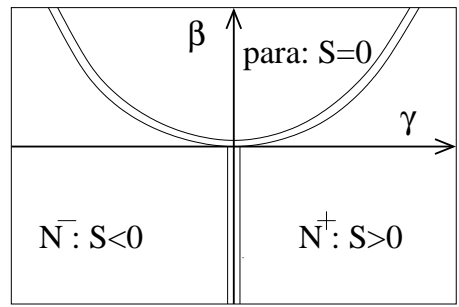} \caption{
Mean field phase diagram for a nematic without anisotropy in the
spin exchange interaction.  The ordered phases N$^\pm$ are
uniaxial. All transitions are first order. }\label{phaseiso}
\end{figure}

We now study the finite temperature phase diagram of the spin
nematic.  We assume that superconductivity is either present as a
background phase throughout the entire region of the phase diagram
that we study, or not present at all. Hence, we do not include
explicitly the interaction between superconducting and nematic
order parameters. As is well known (see, e.g.
Ref~\onlinecite{Chaikin}), symmetry allows the inclusion of cubic
terms into the Ginzburg-Landau (GL) free energy of a nematic order
parameter. Thus, the most general GL free energy for a
spin-isotropic system is, up to quartic order,
\begin{eqnarray}
F_{Q,{\rm iso}}=\beta {\rm Tr}Q_0^2-\gamma {\rm Tr} Q_0^3+\delta
{\rm Tr} Q_0^4. \label{GLisoNem}
\end{eqnarray}
By the identity $({\rm Tr}Q_0^2)^2\equiv 2{\rm Tr}Q_0^4$, which
holds for any $3\times 3$ traceless symmetric matrix, we do not
include the term $({\rm Tr}Q_0^2)^2$ in (\ref{GLisoNem}).  The
phase diagram is shown in Fig.~\ref{phaseiso}.
At high temperatures, $\beta$ is large and positive, and the
system is disordered, $Q_0=0$.  On the other hand, as temperature
is reduced and the value of $\beta$ decreases, there is
spontaneous breaking of spin SU(2) symmetry as the system
undergoes a first order transition into the nematic state. The
nematic in this case is uniaxial as desired (see
sections~\ref{sec:introduction} and \ref{sec:OP}); in order to
stabilize biaxial nematic order, terms of order $Q_0^5$ and
$Q_0^6$ would have to be added to the free energy\cite{Chaikin}.
We note that, as discussed in sections~\ref{sec:OP} and
\ref{sec:waveFn}, $S>0$ in the low temperature phase.  Thus, the
cubic coefficient $\gamma$ must be positive.

How is the above analysis modified by the presence of spin-orbit
effects?  For definiteness, we concentrate on the case of the
cuprate {\LSCO} in the low temperature orthorhombic (LTO) phase.
This compound displays strong evidence for fluctuating stripe
order in its underdoped regime\cite{yamada}, and the spin exchange
constants $J^{\alpha\beta}$ are known in detail from neutron
scattering experiments in undoped {\LCO}\cite{tranquada88,Thio88}.
The analysis presented here can be easily generalized to other
materials. In {\LSCO}, spin-orbit effects are small: The
anisotropic part of the spin exchange interaction
$J^{\alpha\beta}$ is less than $10^{-2}$ of the isotropic part in
undoped {\LCO} \cite{tranquada88,Thio88}. Yet, at low
temperatures, this anisotropy leads to a preferred direction for
the N\'eel order and to weak ferromagnetism \cite{coffey90}. Thus,
although the anisotropy is a low energy effect, playing a weak
role on the onset and magnitude of the nematic order parameter, it
may ultimately fix the preferred spin orientation for the nematic.
This aspect of the interplay between $J$ and $Q$ will be
especially important when we try to separate their contributions
to the anisotropy in the spin susceptibility (see Eq. (\ref{comp})
below).

\begin{figure}
\includegraphics[width=8cm]{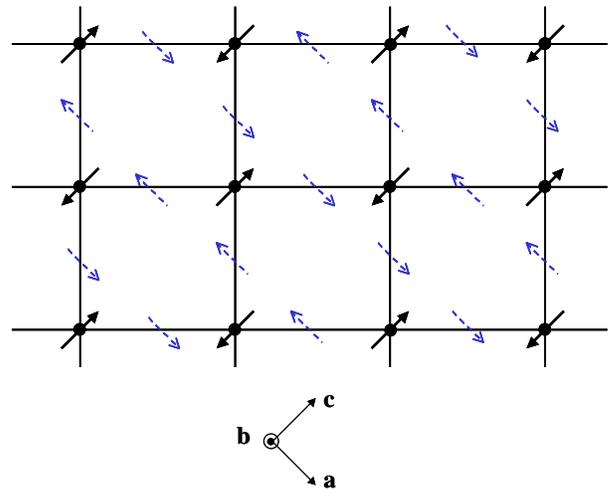} \caption{
The Dzyaloshinskii-Moriya vectors (dashed arrows) and ground state
spin orientation (solid arrows) on a single CuO$_2$ plane in the
LTO phase of {\LCO}.  The black dots denote Cu sites.  The DM
vectors ${\bf D}_{ij}$ point along $\pm{\bf a}$ and are staggered
between adjacent bonds. This, combined with easy plane anisotropy,
yield a weakly canted AF order, where the spins on the Cu sites
have a large staggered component in the $\pm {\bf c}$ direction,
and a small uniform component out of the plane (in the ${\bf b}$
direction, not shown).}\label{fig:DM}
\end{figure}

The most general form of the spin-exchange interaction for spins
on nearest neighbor Cu sites is
\begin{eqnarray}
{\cal H}&=&\sum_{\av{ij}} J_{ij}^{\alpha\beta}S_i^{\alpha}S_j^{\beta}\label{spinexch}\\
&=&\sum_{\av{ij}} \left( J_{0,ij} {\bf S}_i\cdot {\bf
S}_j+J_{s,ij}^{\alpha\beta} S_i^{\alpha} S_j^{\beta}+{\bf
D}_{ij}\cdot ({\bf S}_i\times {\bf S}_j)\right),\nn
%J^{\alpha\beta}=J_0\delta^{\alpha\beta}+\epsilon^{\alpha\beta\gamma}D^{\gamma}+J_{s}^{\alpha\beta},
\end{eqnarray}
where $J_s^{\alpha\beta}$ is a traceless symmetric tensor and
${\bf D}$ is the Dzyaloshinskii-Moriya (DM) vector.  Neither $J_0$
nor $J_s$ depend on the bond $\av{ij}$. It is convenient to work
with the principal axes of the system: the vectors ${\bf a}$ and
${\bf c}$ lie on the CuO$_2$ planes at 45 degrees from nearest
neighbor Cu-Cu bonds, and the vector ${\bf b}$ is normal to the
CuO$_2$ planes, see Fig.~\ref{fig:DM}. In this basis, $J_s$ is
diagonal, $J_0{\bf 1}+J_s={\rm diag}(J^{aa},J^{bb},J^{cc})$ with
$J^{bb}<J^{aa}\approx J^{cc}$ \cite{tranquada88}. For simplicity
we take $J^{aa}=J^{cc}$ in what follows, corresponding to easy
plane antiferromagnetic interactions along the ${\bf a}$-${\bf c}$
plane. On the other hand, the DM vector ${\bf D}_{ij}$ points
along $\pm{\bf a}$, with the sign on each bond $\av{ij}$ given by
the staggered pattern shown in Fig.~\ref{fig:DM}. Thus,
\begin{eqnarray} J_{ij}=\left(\begin{array}{c c c}
J_0+\Delta& 0 & 0 \\
0 & J_0-2\Delta& \pm D \\
0 & \mp D & J_0+\Delta
\end{array}\right),\nn
\end{eqnarray}
where $\Delta=(J^{aa}-J^{bb})/3>0$.  A useful measure of the
relative importance of the anisotropies $\Delta$ and $D$ is given
by the ratio $x\equiv\frac{D^2}{J_0\Delta},$ which is
approximately equal to one in the LTO phase of
{\LCO}\cite{tranquada88,Thio88}.

It is instructive to consider first the effects of spin anisotropy
on an antiferromagnet. The GL free energy $F_N$ of a N\'eel order
parameter ${\bf N}$ is, to lowest order in spin-orbit coupling,
\begin{eqnarray}
F_N=-\lambda {\bf N}^{\rm T} J_s {\bf N}+\frac{\lambda'}{J_0}
({\bf D}\cdot{\bf N})^2+\mu {\bf N}^2+\nu {\bf N}^4.
\label{neelcase}
\end{eqnarray}
The term ${\bf D}\cdot{\bf N}$ is a pseudoscalar and is forbidden
by parity and time-reversal symmetries.  The first two terms
include the effects of anisotropy, and are both quadratic in the
spin-orbit coupling.  The coefficient $\lambda$ is positive, as
necessary to capture the tendency of spins to order along the easy
plane at low temperatures.  Similarly, the DM vector induces a
weak canting for staggered spins that are perpendicular to it.
This pushes ${\bf N}$ towards the plane normal to ${\bf D}$, which
requires $\lambda'>0$.  Thus we see that $J_{\rm s}$ chooses an
easy plane for the spins, and the DM vector selects a preferred
direction, ${\bf c}$, within the easy plane \cite{coffey90}, see
Fig.~\ref{fig:DM}.  The ratio $\lambda'/\lambda$ is unimportant in
this case.

We now turn our attention to the nematic.
%The nematic behaves in a somewhat different manner.
In addition to the usual GL free energy for a nematic order
parameter, eq. (\ref{GLisoNem}), the explicit symmetry breaking
due to anisotropy in the spin exchange can be taken into account,
to quadratic order in the spin-orbit interaction, by adding the
terms,
\begin{eqnarray}
F_Q&=&F_{Q,{\rm iso}}+F_{Q,{\rm anis}} \label{landaufree}\nn\\
F_{Q,{\rm anis}}&=&-\alpha {\rm Tr} (J_{\rm
s}Q_0)+\frac{\alpha'}{J_0} {\bf D}^{\rm T} Q_0 {\bf D}.
\label{spinorbit}
\end{eqnarray}
Note that, unlike the N\'eel case (\ref{neelcase}), the order
parameter $Q$ couples linearly to the anisotropy.  Thus, the
symmetry is broken explicitly, and strictly speaking there is no
disordered phase with $Q_0=0$.  This, however, does not preclude
the existence of crossovers in the order parameter, or even
discontinuities at first order transitions, as we traverse the
phase diagram.  For weak anisotropy the order parameter will be
extremely small at high temperatures, and very sharp crossovers
will be observed. Another consequence of the linear coupling is
that, unlike antiferromagnetic order which always lies on the easy
plane of $J^{\alpha\beta}$, uniaxial nematic order can point
either in the direction of maximal coupling of $J^{\alpha\beta}$,
or in the direction of minimal coupling of $J^{\alpha\beta}$,
depending on the sign of $S$. For the state with nematic order and
$S>0$, we expect the director field ${\bf n}$ to lie on the easy
plane and be orthogonal to ${\bf D}$, {\it i.e.} be parallel to
the direction that the SDW state would take in the absence of
charge fluctuations. This constrains the linear coefficients
appearing in (\ref{spinorbit}) to be positive, $\alpha>0$ and
$\alpha'>0$.

\begin{figure}
\includegraphics[width=6cm]{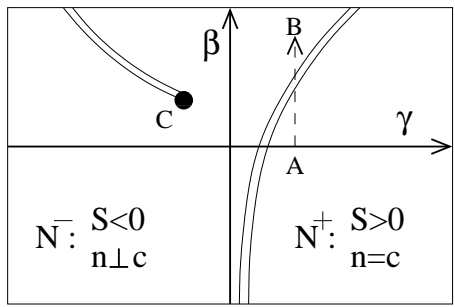} \caption{
Mean field phase diagram including anisotropy in the spin exchange
$J^{\alpha\beta}$, for the case $J^{bb}<J^{aa}=J^{cc}$, ${\bf
D}=\pm{\bf a}$. All transitions are first order.  Phase N$^+$ is
characterized by $S>0$ and ${\bf n}={\bf c}$, phase N$^-$ by
$S<0$.  The director ${\bf n}$ in phase N$^-$ depends on the
parameter $w\equiv\alpha'x/\alpha$: for $w<1$, ${\bf n}={\bf b}$,
while for $w>1$, ${\bf n}={\bf a}$. The parameter $w$ is
material-dependent and is unlikely to change across the phase
diagram.  Thus, only a single type of $N^-$ phase is accessible
within a given material.  The dashed arrow shows a possible
trajectory where a first order transition is crossed as
temperature is increased; alternatively, for smaller values of
$\gamma$, the first order transition may be avoided and strong
crossover behavior may be observed instead.}\label{phasedir}
\end{figure}

\begin{figure}
\includegraphics[width=8cm]{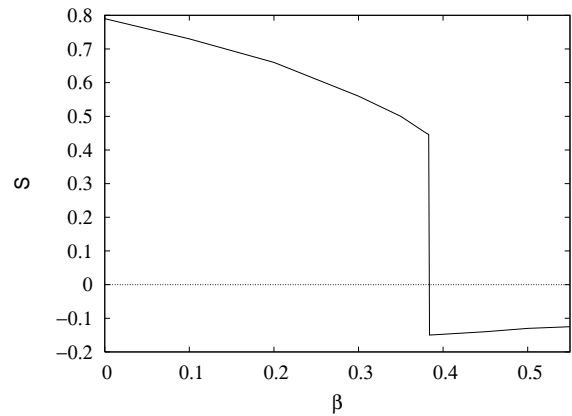} \caption{Value of
$S$ along the dotted trajectory in Fig. \ref{phasedir}. The
parameters $\alpha \Delta=0.1$, $w<1$, and $\gamma=\delta=1$ are
chosen for concreteness; $\beta$ increases with temperature.  As
the anisotropy $\Delta$ is decreased, or for trajectories that
start deep inside the $N^+$ phase due to a larger value of
$\gamma$, the value of $|S|$ in the high temperature phase is
reduced.  Similarly, for small values of $\gamma$, the first order
transition may be avoided and instead a strong crossover within
the $N^-$ phase may be observed.}\label{theta}
\end{figure}

In order to obtain the mean field phase diagram for a spin
nematic, we compare the minima of $F$ for the director field ${\bf
n}$ pointing along the various principal axes $\he{i}$.  The
result is shown in Fig.~\ref{phasedir}.  The low temperature phase
$N^+$ is a uniaxial nematic characterized by $S>0$ and ${\bf
n}={\bf c}$.  The high temperature phase $N^-$ is also uniaxial,
but it has $S<0$.  The director field in $N^-$ depends on the
relative strength of the two anisotropy terms, $w\equiv \alpha'
x/\alpha$. This quantity is material dependent, and is unlikely to
change significantly over the phase diagram (except, of course,
across a structural phase transition), so that only one of the
following scenarios should be observed within a given material:
For $w<1$, ${\bf n}={\bf b}$, whereas for $w>1$, ${\bf n}={\bf
a}$. The coefficient $\beta$ increases with temperature, possibly
tuning a first order phase transition between phases N$^+$ and
N$^-$, or otherwise moving the system through a sharp crossover
within the N$^-$ phase. In either case, the temperature dependence
of the order parameter leads to strong experimental signatures
discussed in Section~\ref{sec:detection}. Figure \ref{theta} shows
the value of $S$ as we move across the N$^+$/N$^-$ phase
transition along the dashed line in Fig. \ref{phasedir}.

\section{Detection}
\label{sec:detection}

In the presence of a spin nematic order parameter
$Q^{\alpha\beta}_{ij}\equiv Q^{\alpha\beta}(\br{i},\br{j})$ of the
form (\ref{nemfact}), symmetry allows the term
\begin{eqnarray}
{\cal H}_{int}=-g\sum_{ij}Q^{\alpha\beta}_{ij}\hat{S}^{\alpha}_i
\hat{S}^{\beta}_j \label{effint}
\end{eqnarray}
to enter the Hamiltonian, which can be thought of as a local
spin-spin interaction mediated by the nematic order. The
coefficient $g$ is positive, as necessary for the stability of the
nematic order.  When the system develops long-range nematic order,
electrons move in a nematic background which acts like an
effective anisotropic spin exchange.  This leads to anisotropy in
the spin response function. In the random phase approximation
(RPA), when an external magnetic field ${\bf B}$ is applied to the
system, ${\cal H}={\cal H}_0+{\cal H}_{int}-\sum_i {\bf B}_i\cdot
{\bf \hat{S}}_i$, the electrons see an effective field
\begin{eqnarray}
B_{\rm eff}^{\alpha}({\bf k})=B^{\alpha}({\bf k}) +g
Q^{\alpha\beta}({\bf k})\langle \hat{S}^{\beta}({\bf k}) \rangle,
\end{eqnarray}
leading to the dynamic spin response
\begin{eqnarray}
\chi^{\alpha\beta}_{\rm RPA}({\bf k},\omega)=\chi_0({\bf
k},\omega) \left[\left(\hat{1}_{2\times2}+g Q({\bf k})\chi_0({\bf
k},\omega)\right)^{-1}\right]^{\alpha\beta} \label{RPA}
\end{eqnarray}
Spin nematic order thus induces anisotropy in the spin
susceptibility.  We introduce a tensor $\Xi^{\alpha\beta}$,
defined by
\begin{eqnarray}
\Xi^{\alpha\beta}({\bf k},\omega)\equiv
\frac{1}{2}\left[(\chi^{-1})^{\alpha\beta}+(\chi^{-1})^{\beta\alpha}\right]-\frac{\delta^{\alpha\beta}}{3}\sum_c(\chi^{-1})^{cc}.\nn
\end{eqnarray}
Experimental measurements of $\chi^{\alpha\beta}$ from polarized
neutron scattering experiments (for arbitrary wave vector ${\bf
k}$) can be used to compute $\Xi^{\alpha\beta}$, which is the
natural object to consider when studying nematic order. In the
current approximation, we find
\begin{eqnarray}
\Xi^{\alpha\beta}_{\rm RPA}({\bf k},\omega)=gQ^{\alpha\beta}({\bf
k}). \label{nemanis}
\end{eqnarray}
Similarly, Knight shift measurements give access to the local
static spin susceptibility, {\it i.e.} the susceptibility
integrated over all wave vectors ${\bf k}$.  To linear order in
the nematic order parameter $Q$, the anisotropy in the Knight
shift over the various principal axes $\alpha$ is proportional to
\begin{eqnarray}
-\int \frac{d^2 k}{(2\pi)^2} \left[\chi_0({\bf
k},\omega=0)\right]^2 \Xi^{\alpha\alpha}({\bf
k},\omega=0),\nonumber
\end{eqnarray}
where no summation over $\alpha$ is implied.

\begin{figure}
\includegraphics[width=5.5cm]{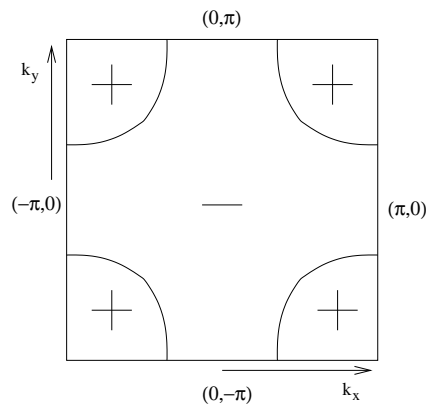}
\caption{Wave vector dependence of $f({\bf k})$ corresponding to
the plaquette nematic operator $\hat{P}_{s_A}$.  This is expected
to be the dominant contribution to $f$, and enters the anisotropy
of the spin susceptibility through Eq. \ref{comp}. The ``+'' and
``-'' regions show the sign of $f({\bf k})$ on the Brillouin zone.
}\label{sym}
\end{figure}

Even in the absence of nematic ordering, spin-orbit coupling leads
to anisotropy in the antiferromagnetic exchange
$J^{\alpha\beta}_{ij}$, which enters the Hamiltonian in a term of
the form (\ref{effint}) with $g Q\to -J$, see
eq.~(\ref{spinexch}).  The complete expression for
$\Xi^{\alpha\beta}$ is therefore
\begin{eqnarray}
\Xi^{\alpha\beta}_{\rm RPA}({\bf k},\omega)= g
Q^{\alpha\beta}({\bf k})-J_{\rm s}^{\alpha\beta}({\bf k}).
\label{comp}
\end{eqnarray}
Note that only the traceless symmetric part of $J_{ij}$
contributes to $\Xi$.  We face the challenge of untangling the
contributions to (\ref{comp}) due to the anisotropy in $J$ from
those coming from the presence of nematic order.  For this, the
analysis of Section~\ref{sec:waveFn} is pivotal, in particular eq.
(\ref{domform}), which gives the dominant wave vector dependence
of the nematic contribution to the anisotropy in the spin response
function, see Figure~\ref{sym}. This can be combined with detailed
knowledge of the form of $J_{\rm s}$ \cite{tranquada88,Thio88},
$J_{{\rm s}}({\bf k})=\eta({\bf k})\Delta\,{\rm diag}(1,1,-2)$. As
before, $\Delta=(J^{aa}-J^{bb})/3>0,$ and we assume that
anisotropy is small beyond the nearest-neighbor range to set
$\eta({\bf k})=2(\cos k_x+\cos k_y)$.  Thus, the contributions due
to $Q$ and $J$ can be distinguished by the wave vector dependence
of the signal, as one is proportional to $f({\bf
k})=\frac{4}{\sqrt{6}}(\cos k_x \cos k_y -\cos k_x -\cos k_y)$ and
the other to $\eta({\bf k})=2(\cos k_x+\cos k_y)$. For example,
experiments measuring the susceptibility at the points ${\bf
k}=(\pi,0)$ and $(0,\pi)$ are sensitive to $Q$, but not $J$, see
Fig. \ref{sym}.

%We examine two possible ways of doing so, comparing
%the spatial symmetries ({\it i.e.} wave vector dependencies)
%of the anisotropies generated by these two mechanisms, and
%looking for non-analyticity in the anisotropy due to phase transitions in the nematic order.
For completeness, we evaluate the expression (\ref{comp}) in the
various phases shown in Fig.~\ref{phasedir} for underdoped
{\LSCO}. In the low temperature phase N$^+$, we expect,
\begin{eqnarray}
\Xi^{aa}&=&-g|S|f({\bf k})/3-\Delta\eta({\bf k}),\nonumber\\
\Xi^{bb}&=&-g|S|f({\bf k})/3+2\Delta\eta({\bf k}),\nonumber\\
\Xi^{cc}&=&2g|S|f({\bf k})/3-\Delta\eta({\bf k}),\nonumber
\end{eqnarray}
In the high temperature phase N$^-$, we must consider two separate
scenarios. If $w>1$, then ${\bf n}={\bf a}$, and
\begin{eqnarray}
\Xi^{aa}&=&-2g|S|f({\bf k})/3-\Delta\eta({\bf k}),\nonumber\\
\Xi^{bb}&=&g|S|f({\bf k})/3+2\Delta\eta({\bf k}),\nonumber\\
\Xi^{cc}&=&g |S|f({\bf k})/3-\Delta\eta({\bf k}),\nonumber
\end{eqnarray}
whereas, if $w<1$, then ${\bf n}={\bf b}$, and
\begin{eqnarray}
\Xi^{aa}&=&-\Xi^{bb}/2=\Xi^{cc},\nn\\
\Xi^{cc}&=&g|S|f({\bf k})/3-\Delta\eta({\bf k}),\nonumber
\end{eqnarray}
Thus, if the system goes through a phase transition between phases
N$^+$ and N$^-$, the discontinuity in $S$, shown in
Fig.~\ref{theta}, together with the change in the director field
${\bf n}$, would give a very clear experimental signature of
nematic order in Knight shift and polarized neutron scattering
experiments.
%Note that, if
%$\eta>1$, then the nematic contribution in the high temperature phase
%$N^-$ can be isolated by taking the difference
%\begin{eqnarray}
%\Xi^{aa}-\Xi^{cc}=g|S|f({\bf k}).
%\end{eqnarray}

Finally, we consider the prospect of observing director density
waves (DDW), the Goldstone modes corresponding to the nematic
state, in neutron scattering experiments.  To do this, we study
the pole structure of the RPA result (\ref{RPA}). Expanding
(\ref{RPA}) about ${\bf k}=0$ and $\omega=0$, we find two
degenerate DDW modes, with speed $v_{\rm DDW}\propto\sqrt{g|S|}$,
\begin{eqnarray}
\chi^{\perp\perp}({\bf k},\omega)\propto \frac{{\bf
k}^2}{\omega^2-v_{\rm DDW}^2 {\bf k}^2}, \label{rpaDDW}
\end{eqnarray}
with $\perp$ labelling either of the directions perpendicular to
the director of the nematic. Equation (\ref{rpaDDW}) is consistent
with an independent calculation of the DDW modes starting from an
effective low energy quantum rotor model, see
Section~\ref{sec:rotor}. We note that the DDW modes found here may
be overdamped, depending on the details of the system. This would
be the case, for instance, if $v_{\rm DDW}$ were smaller than the
Fermi velocity. However, provided that the modes are underdamped,
they can be detected in neutron scattering experiments.  This is
surprising at first, as local $Z_2$ invariance implies that
equal-time correlators of the from $\langle S^{\alpha}_i
S^{\beta}_j\rangle$ must decay exponentially at large distances
$|{\bf x}_i-{\bf x}_j|$. However, as (\ref{rpaDDW}) indicates, the
correlator at different times need not decay exponentially. Also
note that the inelastic scattering peak (\ref{rpaDDW}) has
vanishing weight as ${\bf k}\to 0$, consistent with the absence of
an elastic Bragg peak. Unfortunately, this feature makes DDW modes
difficult to distinguish from the low energy collective modes of a
quantum paramagnet, computed in Appendix~\ref{App:paramagnet}.

%However, DDW and quantum paramagnons behave differently in a weak
%magnetic field. By symmetry, if an applied field is parallel to
%the director, whose direction is pinned by microscopic anisotropy,
%both DDW modes remain degenerate. As the field is rotated relative
%to the director field, the degeneracy is lifted.  On the other
%hand, the low energy modes are always be split by a magnetic
%field, and the spectrum is independent of the direction of the
%magnetic field.

\section{Quantum rotor model}
\label{sec:rotor}

Another approach to study Goldstone modes is to introduce a
quantum rotor model which captures the low energy properties of
the system. This can be done by coarse-graining spins in
plaquettes, as carried out in section~\ref{sec:waveFn}. There we
found that the low energy plaquette states at half-filling contain
one singlet ground state, and one triplet state obtained by acting
on the ground state by the staggered magnetization operator.  This
matches the low energy spectrum of a quantum rotor ${\bf
\hat{n}}_i$ on the plaquette $i$, which can be written in terms of
the component spins of the plaquette as
\begin{eqnarray}
\hat{\bf L}_i&=&\hat{\bf S}_{i1}+\hat{\bf S}_{i2}
+\hat{\bf S}_{i3}+\hat{\bf S}_{i4},\nonumber\\
\hat{\bf n}_i&\sim&\hat{\bf S}_{i1}-\hat{\bf S}_{i2} +\hat{\bf
S}_{i3}-\hat{\bf S}_{i4}.\nonumber
\end{eqnarray}
Here, the total spin $\hat{\bf L}$ acts as the canonical conjugate
of $\hat{\bf n}$, satisfying the commutation relations,
\begin{eqnarray}
\left[ L_i^{\alpha},L_j^{\beta}\right]&=&i\delta_{ij}\epsilon^{\alpha\beta\gamma}L_i^{\gamma},\nn\\
\left[ L_i^{\alpha},n_j^{\beta}\right]&=&i\delta_{ij}\epsilon^{\alpha\beta\gamma}n_i^{\gamma},\nn\\
\left[ n_i^{\alpha},n_j^{\beta}\right]&=&0.
\end{eqnarray}
We introduce an effective Hamiltonian for these rotors, consistent
with local $Z_2$ symmetry,
\begin{eqnarray}
H_{\rm rotor}=\frac{\nu}{2}\sum_i {\hat{\bf L}}_i^2-
\tilde{g}\sum_{\langle ij\rangle}(\hat{\bf n}_i\cdot\hat{\bf
n}_j)^2.\label{rotorHam}
\end{eqnarray}
The dynamic spin susceptibility can be obtained, in the long wave
length limit ${\bf k}\to 0$, by the linear response of ${\bf L}$
to an external field ${\bf B}_i={\bf B}e^{-i(\omega t-{\bf k}\cdot
{\bf r}_i)}$,
\begin{eqnarray}
H_{\bf B}^0=-\sum_i {\bf B}_i\cdot \hat{\bf L}_i.
\end{eqnarray}
The Heisenberg equations of motion for the rotors become
\begin{eqnarray}
\frac{d\hat{\bf L}_i}{dt}&=&\tilde{g}\sum_{j\in NN(i)}(\hat{\bf
n}_i\cdot\hat{\bf n}_j)(\hat{\bf n}_i\times\hat{\bf n}_j)
+{\bf B}_i\times\hat{\bf L}_i\nn\\
\frac{d\hat{\bf n}_i}{dt}&=&\frac{\nu}{2}(\hat{\bf
L}_i\times\hat{\bf n}_i-\hat{\bf n}_i\times\hat{\bf L}_i) -{\bf
B}_i\times\hat{\bf n}_i \label{EOM}
\end{eqnarray}
In the absence of an external field ${\bf B}$, the rotor
correlator is assumed to be highly local,
\begin{eqnarray}
\langle \hat{n}_i^{\alpha} \hat{n}_j^{\beta} \rangle_0=
\delta_{ij}(\delta^{\alpha\beta} n^2/3+Q^{\alpha\beta})\equiv
\delta_{ij} G^{\alpha\beta}.\nonumber
\end{eqnarray}
The product of $\hat{n}$ operators then differs from the mean
field result by a correction,
\begin{eqnarray}
\hat{n}_i^{\alpha} \hat{n}_j^{\beta}-%\langle \hat{n}_i^{\alpha} \hat{n}_j^{\beta}\rangle_0
\delta_{ij} G^{\alpha\beta}%
=\delta_{ij}\rho^{\alpha\beta}+\lambda_{ij}^{\alpha\beta},\nonumber
\end{eqnarray}
which has been split into a local term $\rho^{\alpha\beta}$, and a
non-local term, $\lambda_{ij}^{\alpha\beta}=0$ for $i=j$.
Linearizing the equations of motion (\ref{EOM}) with respect to
these corrections, we obtain
\begin{eqnarray}
-i\omega L^{\alpha}&=&F({\bf k})\epsilon^{\alpha\beta\gamma}G^{\beta\delta}\rho^{\gamma\delta},\nn\\
-i\omega\rho^{\alpha\beta}&=&(\epsilon^{\beta\gamma\delta}G^{\alpha\delta}+\epsilon^{\alpha\gamma\delta}G^{bd})(\nu
L^{\gamma}-B^{\gamma}) \label{mfEOM}
\end{eqnarray}
where
\begin{eqnarray}
F({\bf k})=2 \tilde{g}\sum_{i=1}^d(1-\cos q_i).
\end{eqnarray}
Since we are interested in the linear response, we have excluded
from Eq.~(\ref{mfEOM}) terms quadratic in ${\bf B}$.  Note that,
to this order in ${\bf B}$, the non-local term $\lambda_{ij}$
drops out of the equations of motion.

For the current case of interest, a uniaxial nematic,
$G^{\alpha\beta}$ is of the form
\begin{eqnarray}
G=\frac{1}{3}\left(\begin{array}{c c c}
n^2-S & 0 & 0 \\
0 & n^2-S & 0 \\
0 & 0 & n^2+2S
\end{array}\right).\nonumber
\end{eqnarray}
Solving (\ref{mfEOM}) for ${\bf L}$ yields
\begin{eqnarray}
L^\perp&=&-\frac{S F({\bf k}^2)}{\omega^2-\nu S^2 F({\bf
k})}B^\perp,
\label{mfDDW}\\
L^{\hat{n}}&=&0,\nonumber
\end{eqnarray}
where $\perp$ is either of the directions perpendicular to
$\hat{\bf n}$. Note that, unlike the case of antiferromagnetic
order, $\chi^{\alpha\beta}$ has no off-diagonal contributions.
This is attributed to time reversal invariance. For long wave
lengths ${\bf k}\to 0$, $F({\bf k})\to\tilde{g}{\bf k}^2$, and
poles in the susceptibility indicate the presence of two
degenerate DDW modes, with velocity $v_{\rm DDW}=\sqrt{\nu S^2
\tilde{g}}$. This is consistent with the RPA result found above,
Eq. (\ref{rpaDDW}).

Finally, we compute the dynamic spin susceptibility near
$k=(\pi,\pi)$ by replacing the term $H_{\bf B}^0$ by a source that
couples directly to the staggered magnetization,
\begin{eqnarray}
H_{\bf B}^\pi=-\sum_i {\bf B}_i\cdot {\bf \hat{n}}_i.
\end{eqnarray}
Proceeding as above, we find the linear response
\begin{eqnarray}
n_i=\frac{-\nu G^{\alpha\beta}}{\omega^2-z\tilde{g}
n^2}B_i^{\beta},
\end{eqnarray}
where $z=4$ is the coordination number of the lattice. Note that,
in the current approximation, the collective modes near
$(\pi,\pi)$ are non-dispersing and gapped. The spin gap
$\omega=\sqrt{z\tilde{g} n^2}$ indicates that there is no long
range N\'eel order in the system.

\section{Lattice gauge theory}
\label{sec:LGT}

\begin{figure}
\includegraphics[width=8cm]{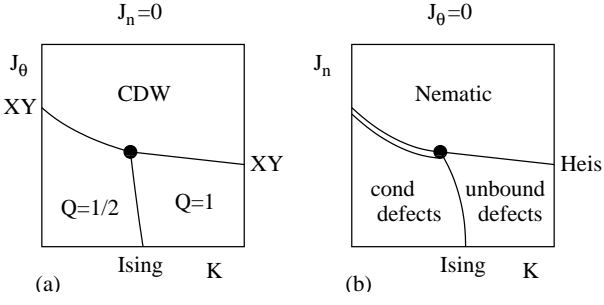}
\caption{ (a) Phase diagram for eq. (\ref{fracaction}) for the
case $J_n=0$, corresponding to a $Z_2$ lattice gauge theory with a
$U(1)$ field. The Ising transition as $K$ is increased corresponds
to a binding of $Q=1/2$ topological excitations into $Q=1$
excitations. (b) Case $J_\theta=0$, corresponding to a $Z_2$
lattice gauge theory with an $SO(3)$ field.  The first order
nematic to paramagnet transition can be split into two second
order transitions, passing through a topologically ordered phase.
}\label{frac2}
\end{figure}

\begin{figure}
\includegraphics[width=8cm]{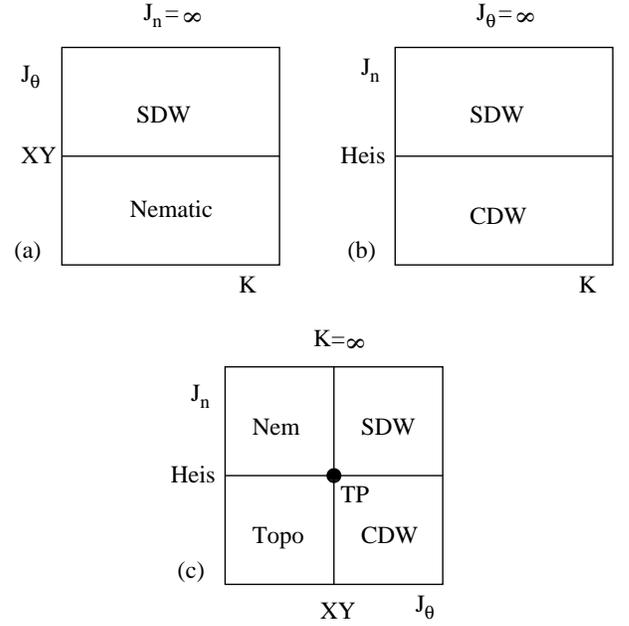}
\caption{ Phase diagrams when any of the couplings in eq.
(\ref{fracaction}) is infinite.  In all of these cases,
topological defects in the gauge field are energetically
forbidden, and we can work in minimal gauge $\sigma_{ij}=1$. (a)
Case $J_n=\infty$, insuring that at least nematic order is
present.  $K$ is irrelevant in this situation, as the field
$\sigma_{ij}$ is frozen.  As $J_\theta$ is increased, the system
undergoes an XY transition from nematic to SDW order.  The
situation is similar to the case $J_\theta=\infty$, shown in (b),
but now a Heisenberg transition is seen from CDW to SDW order.  In
(c), $K=\infty$, leading to a dynamical decoupling of the $\theta$
and ${\bf n}$ terms in the action (\ref{fracaction}). For small
$J_\theta$ and $J_{\bf n}$, a paramagnet with only topological
order survives. }\label{frac1}
\end{figure}

Up to now, we have ignored the possibility of fractionalization
and the richer phase structure that it allows.  As an example, by
introducing a disclination core energy, it is possible to split
the phase transition for the onset of nematic order, which is
strongly first order in Landau theory, into two second order
transitions \cite{lammert93}. For a theory with local $Z_2$ gauge
redundancy, it is useful to discretize the magnetic degrees of
freedom on a lattice and to introduce an auxiliary gauge field
$\sigma_{ij}=\pm 1$ living on the bonds of the lattice. Under a
gauge transformation at site $i$, the site variables pick up a
minus sign, ${\bf n}_i\to-{\bf n}_i$ and $e^{i\theta_i}\to
-e^{i\theta_i}$, while simultaneously the bond variables
surrounding $i$ change sign, $\sigma_{ij}\to-\sigma_{ij}$.  The
simplest gauge invariant action that can be written under these
conditions is
\begin{eqnarray}
S=-J_n\sum_{\langle ij\rangle}{\bf n}_i\sigma_{ij}{\bf n}_j
  &-&J_\theta\sum_{\langle ij\rangle}\sigma_{ij}\cos(\theta_i-\theta_j)\nonumber\\
  &-&K\sum_{\Box}\prod_{\Box}\sigma_{ij}.
\label{fracaction}
\end{eqnarray}
Unlike section~\ref{sec:meanField}, here we ignore the anisotropy
in the tensor $J^{\alpha\beta}$, which is weak and only affects
the very low energy physics. Figures \ref{frac2} and \ref{frac1}
display the phase diagrams for extreme values of the couplings
$J_n$, $J_\theta$ and $K$. Figure \ref{frac2} shows the phase
diagrams for the cases $J_n=0$ and $J_\theta=0$.  These correspond
to $Z_2$ lattice gauge theories with $U(1)$ and $SO(3)$ Higgs
fields, respectively, which have been studied extensively in the
literature\cite{lammert93,scalapino02,Senthil00,DemlerFract}.

Figure \ref{frac1} shows the situation when any of the couplings
in (\ref{fracaction}) is infinite.  In all of these cases, the
auxiliary field $\sigma_{ij}$ is completely ordered, and we can
work in ``minimal gauge'', where $\sigma_{ij}=1$ for all bonds.
This corresponds to a lack of topological defects, which are
energetically forbidden.  Once a value of ${\bf n}$ (or $\theta$)
is chosen at a given lattice point, smoothness of the fields
insures that the $Z_2$ redundancy in ${\bf n}$ ($\theta$) is
removed everywhere.  Thus, the universality class of these
transitions is the same as that of the corresponding ungauged
theories. For instance, in Fig. \ref{frac1}(c), starting with the
topologically ordered paramagnet at small $J_n$ and $J_\theta$, we
can increase $J_\theta$ until the onset of CDW order through an XY
phase transition.  Similarly, by increasing $J_n$ we get the onset
of nematic order through a Heisenberg transition. In these regions
of phase space, the simultaneous presence of both CDW and nematic
orders imply SDW order.  To see this, suppose that $\langle {\bf
n} \rangle_m$ and $\langle e^{i\theta} \rangle_m$ are
simultaneously non-zero, where the subscript $m$ indicates that we
are working in minimal gauge. Then $\langle e^{i\theta}{\bf
n}\rangle_m$ is non-zero, but this quantity is in fact independent
of the gauge used.

Let us now consider the phase diagram when $K=0$.  In this case
the auxiliary gauge field fluctuates strongly, and it is useful to
sum (\ref{fracaction}) over $\sigma_{ij}$ configurations to get
\begin{eqnarray}
S_{\rm eff}&=&-\frac{J_n^2}{2}\sum_{\langle ij \rangle}({\bf
n}_i-{\bf n}_j)^2
-\frac{J_\theta^2}{2}\sum_{\langle ij \rangle}\cos^2(\theta_i-\theta_j) \nonumber\\
&-&J'\sum_{\langle ij \rangle}{\bf n}_i\cdot{\bf n}_j
\cos(\theta_i-\theta_j)+\ldots \label{seff}
\end{eqnarray}
In this particular model, $J'=J_n J_\theta$.  An analysis of two
dimensional algebraic order in a model of this sort (with XY
spins) has been carried out in Ref.~\onlinecite{kruger02}.  There,
the only phases present are paramagnet, CDW, nematic, and SDW, and
there is a direct transition between paramagnetic and SDW phases.
Thus, for the action (\ref{fracaction}), although we only look at
the boundary of the phase diagram , in all the limits considered,
the simultaneous appearance of nematic and CDW order implies SDW
order.

\begin{figure}
\includegraphics[width=8cm]{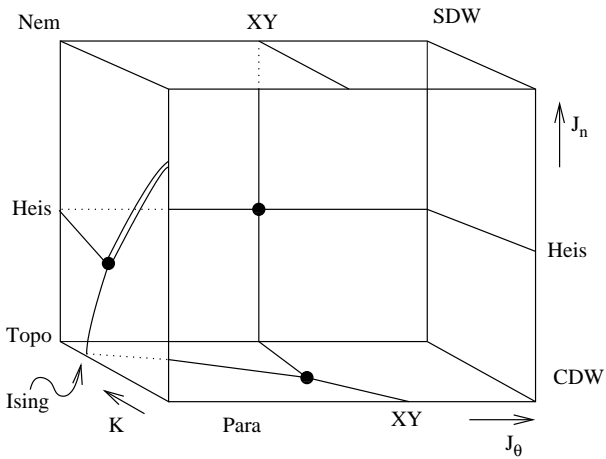}
 \caption{
Five faces of the phase diagram of the action (\ref{fracaction}):
$J_\theta=0$, $J_\theta=\infty$, $J_n=0$, $J_n=\infty$, and
$K=\infty$.  We do not compute the phase diagram along the sixth
face, $K=0$, explicitly.  However, there are indications from XY
spins in 2d that the only phases present there are paramagnet,
CDW, SDW, and nematic\cite{kruger02}. However, introduction of a
new term (\ref{extraterm}) can spit the SDW into a CDW+nematic
phase.  Note that, for large values of $K$, a topologically
ordered phase becomes available.}\label{frac3}
\end{figure}

%\begin{figure}
%\includegraphics[width=8cm]{figures/frac4}
% \caption{
%The slice $K=0$.  The largely fluctuating gauge field is
%integrated out to yield an effective theory for the spin and
%charge fields; the paramagnetic phase is confining.  {\bf These
%results are for classical XY spins in two dimensions
%\cite{kruger02}.  Thus, they describe quasi long range order, and
%the first order transition shown is actually second order.}
%}\label{frac4}
%\end{figure}

However, it is possible to have a state with nematic and CDW
orders without simultaneously stabilizing SDW order.  To see this,
notice that $J'$ in (\ref{seff}) can be tuned by adding an extra
gauge invariant term to the original action (\ref{fracaction})
\begin{eqnarray}
\Delta S=-J_e\sum_{\langle ij \rangle}{\bf n}_i\cdot{\bf n}_j
\cos(\theta_i-\theta_j). \label{extraterm}
\end{eqnarray}
In particular, when $J'=0$, Eq. (\ref{seff}) leads to two
independent transitions for ${\bf n}$ and $\theta$, yielding
paramagnetic, CDW, nematic, and nematic+CDW phases.  The latter
phase breaks SU(2) spin symmetry and translational symmetry, but
leaves time reversal invariant.  How can such a state arise from
fluctuations of a spin stripe?  In the stripe picture, an
anti-phase domain wall serves a dual role, both as a region where
charge accumulates and as a boundary between domains of opposite
staggered magnetization.  The effect of (\ref{extraterm}) is to
disentangle these two roles to obtain two separate objects, a
charge line component and a magnetic anti-phase line component.
Then, fluctuations of the anti-phase line component can restore
time reversal symmetry without restoring translational invariance.
Note that, as the extra coupling $J_e$ is decreased, an Ising
phase transition between nematic+CDW and SDW phases begins to
occur in the corner of the phase diagram $J_\theta,J_n\to\infty$
until, for small enough $J_e$, the nematic+CDW phase disappears
entirely in favor of the SDW phase.

\section{Summary}

In this paper we have considered the prospect of direct detection
of a quantum spin nematic in strongly correlated electron systems,
such as heavy-fermion compounds, the cuprates, and the organic
superconductors.  The spin nematic order parameter is a spin-2
operator, and it does not couple, to linear order, to many of the
conventional probes, such as neutrons, photons, or nuclear spins.
However, we show that electrons moving in a nematic background
have an anisotropic spin susceptibility, which can be detected in
polarized neutron scattering and Knight shift experiments.  In
addition, we discuss the possibility of observing the Goldstone
modes associated with nematic ordering in inelastic neutron
scattering experiments.

In Section~\ref{sec:OP}, we defined the nematic order parameter in
terms of the equal-time spin-correlation function, and argued for
uniaxial nematic order in systems with collinear spin
correlations. In Section~\ref{sec:waveFn}, we introduced wave
functions to describe nematic order in a number of situations,
including a stand-alone nematic, and nematic order coexisting
either with spin-singlet or spin-triplet superconductivity.  In
order to do this, we followed a coarse-graining and low energy
projection procedure, as done in Refs.~\onlinecite{Zhang1999}
and~\onlinecite{Altman2002}.  We used these wave functions to
constrain the short distance and low energy structure of the
nematic order parameter. In Section~\ref{sec:meanField}, we
considered the finite temperature phase diagram of the nematic
through a mean-field analysis of the GL free energy.  Here, we
included the effects of the spin-orbit interaction which, even
when weak, ultimately fix the director of the nematic.   We
illustrated this principle using spin-exchange constants measured
in undoped {\LCO}. In Section~\ref{sec:detection}, we computed the
anisotropic spin response of electrons in the nematic phase, and
discussed prospects for its observation.  There, we also computed
the spectrum of Goldstone modes of a quantum spin nematic. In
Section~\ref{sec:rotor}, these results were supported by the
analysis of an effective quantum rotor model.  Finally, in
Section~\ref{sec:LGT}, we considered the possible presence of
topologically-ordered phases, as well as an exotic CDW+nematic
phase, in a system of fractionalized electrons.

Before concluding, we would like to mention a few experimental
systems where nematic order may be found (in addition to \LSCO,
which is discussed in the introduction
\cite{zaanenNussinov,nussinovZaanen}).  In particular, heat
capacity measurements in two-dimensional solid $^3$He support the
presence of large many-spin interactions \cite{ishida}. On the
other hand, numerical simulations of spins on a square lattice
with ring exchange indicate a stable ``p-nematic" phase
\cite{lauchli}.  In this case, frustration due to the ring
exchange can destroy antiferromagnetic order in favor of a nematic
phase.  Another experimental system of interest is V$_2$O$_3$
which, as temperature and pressure are varied, displays a Mott
transition along a first order line ending in a second order
critical point \cite{limelette}.  The lattice constants change
discontinuously across the first order line, leading to a change
in the ratio of potential to kinetic energies, and thus to a
metal-insulator transition.  The topology of the phase diagram is
the same as that shown in the N$^-$ phase surrounding point C in
Fig.~\ref{phasedir}.  In fact, as discussed in
section~\ref{sec:meanField}, spin nematic ordering couples
linearly to the anisotropy in the spin exchange.  Thus,
discontinuity in the spin nematic order parameter leads to a
discontinuous deformation of the lattice, which could explain the
phase diagram of V$_2$O$_3$.

We thank E. Altman, A. Auerbach, B. Halperin, M. Hastings, J.-P.
Hu, A. Imambekov, C. Nayak, A. Paramekanti, S. Sachdev, A.
Vishwanath, and J. Zaanen for useful discussions. This work was
supported by Harvard NSEC, NSF grant No. DMR-01-32874.

%\section{To do:}
%\begin{enumerate}
%\item Understand better Fig. \ref{frac3}.
%\item Add references.
%\item Conclusion.
%\end{enumerate}

\appendix

\section{Correlations of spin and charge}
\label{App:spinCharge}

While equation (\ref{nonfact}) gives the simplest wave function
that describes a state with both superconducting and nematic
orders, it does not incorporate correlations between charge and
spin degrees of freedom.  In the stripe picture, domain walls are
not only regions where holes accumulate, they are also regions
across which the staggered magnetization changes sign.  One can
modify (\ref{nonfact}) to include this effect, by relating
$\sigma_i$ to the number of domain walls crossed:
\begin{eqnarray}
|\psi_3\rangle&=&\sum_{\{ \sigma_1,\sigma_2\ldots \}=\pm1}\prod_k
\delta\left( \sigma_k-\sigma_1(-1)^{\sum_{j=1}^k
\hat{b}^\dagger_j\hat{b}_j}\right)
g\left[\hat{b}^\dagger_j\hat{b}_j\right]\nn\\
&\,& \times\left[\prod_i(s+\sigma_i
m^{\alpha}\tdi+c\bd_i)\Om_i\right]. \label{wave1}
\end{eqnarray}
In this case, the sum over sites inside the delta function runs
over a path on the plane joining site 1 to site $k$.  This forces
a change of sign in $\sigma_k$ every time a new domain wall
intervenes between those two sites.  The path sum is only
independent of path whenever the domain walls run continuously
through the sample instead of ending up abruptly, and when an odd
number of them do not intersect, as would happen, for instance, in
a ``T-junction''. The functional $g$ is chosen to destroy all
configurations that violate these constraints, thus enforcing the
integrity of the fluctuating domain walls.  The simplest choice
for $g$ is a projection operator that gives equal weight to all
allowed configurations.  In this case, the wave function
(\ref{wave1}) can be rewritten as
\begin{eqnarray}
|\psi_4\rangle=\sum_{\cal D}\prod_{i\not\in {\cal
D}}(s+\sigma_{{\cal D},i} m^{\alpha}\tdi)\Om_i \prod_{j\in
\cal{D}} c\bd_j \Om_j, \label{wave2}
\end{eqnarray}
where ${\cal D}$ denotes an allowed domain-wall configuration, and
$\sigma_{{\cal D},i}$ is the sign of the staggered magnetization
on site $i$ in configuration ${\cal D}$.
%  Alternatively, instead of weighting
%the wave function by $g$, one can enforce the integrity of the domain walls by taking
%a product of delta functions over all possible paths joining sites 1 and $k$,
%\begin{eqnarray}
%\sum_{\{ \sigma_1,\sigma_2\ldots \}=\pm1}\prod_k\left[\prod_{{\rm paths } P_{1k}}
%\delta\left( \sigma_k-\sigma_1(-1)^{\sum_{P_{1k}} \hat{b}^\dagger_j\hat{b}_j}\right)\right]
%\label{wave2}\\
%\left[\prod_i(s+\sigma_i m^{\alpha}\tdi+c\bd_i)\Om_i\right].
%\nonumber
%\end{eqnarray}
Equations (\ref{wave1}) and (\ref{wave2}) are alternate
descriptions of the nematic state of Zaanen and Nussinov
\cite{zaanenNussinov,nussinovZaanen}.

\section{Collective modes of a quantum paramagnet}
\label{App:paramagnet}

We would like to see if the director density waves (DDW) of a
nematic state can be easily distinguished from the collective
modes of a quantum paramagnet.  These latter modes can be derived,
in the RPA approximation, by substituting
$-gQ^{\alpha\beta}_{ij}\to J_0\delta^{\alpha\beta}\eta_{ij}$ in
equations (\ref{effint}) and (\ref{RPA}).  Here, $\eta_{ij}$ is
equal to 1 if $i$ and $j$ are nearest neighbors, and zero
otherwise; and $J_0>0$ is the antiferromagnetic spin exchange. We
find an isotropic spin response,
\begin{eqnarray}
\chi_{\rm RPA}^{\alpha\beta}({\bf k},\omega)=\frac{\chi_0({\bf
k},\omega)}{1-J_0 \eta({\bf k})\chi_0({\bf
k},\omega)}\delta^{\alpha\beta},\label{isotropicResp}
\end{eqnarray}
where $\eta({\bf k})=2(\cos k_x+\cos k_y)$.  In the limit ${\bf
k}\to 0$, $\omega\to 0$, the free spin susceptibility tends to
$\chi_0\to k^2/\omega^2$, leading to the long wavelength result,
\begin{eqnarray}
\chi_{\rm RPA}^{\alpha\beta}%({\bf k}\approx 0,\omega\approx 0)
\sim \frac{k^2}{\omega^2-4 J_0 k^2}\delta^{\alpha\beta}.\nn
\end{eqnarray}
Thus, there are three degenerate gapless paramagnon poles (with
vanishing weight at ${\bf k}=0$).
%  As discussed in
%Section~\ref{sec:detection}, magnetic field experiments can be
%used to distinguish between paramagnons and DDW modes.
On the other hand, the poles of Eq.~(\ref{isotropicResp}) near
${\bf k}=(\pi,\pi)$ have a spin gap $\Delta$ due to the lack of
long-range N\'eel order,
\begin{eqnarray}
\omega_{{\bf k}\sim(\pi,\pi)}=\sqrt{4J_0^2 \left({\bf
k}-(\pi,\pi)\right)^2+\Delta^2}.\nn
\end{eqnarray}

%Note the contrast between the DDW modes derived in
%Sections~\ref{sec:detection} and~\ref{sec:rotor}, and the
%excitations described here. Only two DDW modes exist, one for each
%direction perpendicular to the director $\hat{\bf n}$ of the
%nematic, instead of the three antiparamagnon modes.  The DDW modes
%are gapless, and they are centered at ${\bf k}=0$, whereas the
%antiparamagnons are gapped, and centered at ${\bf k}=(\pi,\pi)$.

%\bibliography{D:/Research/Thesis/bibs}
%\end{document}

\end{document}